\documentclass[aps,pre,twocolumn,groupedaddress]{revtex4-1}

\usepackage{color}

\usepackage[utf8]{inputenc}
\usepackage[english]{babel}
\usepackage{amsmath,graphicx,enumerate}

\begin{document}


\title{Role of rotational inertia for collective phenomena in active matter}

\author{Lorenzo Caprini$^{1}$*}
\author{Rahul Kumar Gupta$^{1}$}
\author{Hartmut L\"owen$^{1}$}
\affiliation{$^1$ Heinrich-Heine Universit\"at D\"usseldorf, D\"usseldorf, Germany. }

\email{lorenzo.caprini@gssi.it}

\date{\today}


\begin{abstract}

We investigate the effect of rotational inertia on the collective phenomena of underdamped active systems and show that the increase of the moment of inertia of each particle favors non-equilibrium phase coexistence, known as motility induced phase separation, and counteracts its suppression due to translational inertia.
Our conclusion is supported by a non-equilibrium phase diagram (in the plane spanned by rotational inertial time and translational inertial time) whose transition line is understood theoretically through scaling arguments.
In addition, rotational inertia increases the correlation length of the spatial velocity correlations in the dense cluster. 
The fact that rotational inertia enhances collective phenomena, such as motility induced phase separation and spatial velocity correlations, is strongly linked to the increase of rotational persistence.
Moreover, large moments of inertia induce non-monotonic temporal (cross) correlations between translational and rotational degrees of freedom truly absent in non-equilibrium systems.

\end{abstract}

\maketitle

\section{Introduction}


Physical or biological systems of active particles~\cite{marchetti2013hydrodynamics, elgeti2015physics, bechinger2016active, gompper20202020} are not only common at the microscopic but also at the macroscopic scales~\cite{klotsa2019above}.
Typical examples in the animal world are birds~\cite{cavagna2014bird}, showing flocking~\cite{cavagna2010scale}, fish, displaying schooling~\cite{pavlov2000patterns}, as well as penguins~\cite{zampetaki2021collective} or flying beetles~\cite{mukundarajan2016surface}, giving rise to a broad range of fascinating collective phenomena.
In addition, inanimate objects such as walking droplets~\cite{valani2019superwalking} or flying whirling fruits~\cite{rabault2019curving} represent other examples of macroscopic self-propelled particles.
Recently, active granular systems~\cite{weber2013long, walsh2017noise, dauchot2019dynamics, kumar2019trapping, leoni2020surfing, gupta2022active}, self-propelling because of some asymmetry in their shapes, have been investigated as a prototype of biological active matter as well as for their broad range of applications, for instance in the design of robots~\cite{leyman2018tuning} even able to self-organize~\cite{scholz2018rotating}.

In this broad class of self-propelling macroscopic systems, inertial effects play a pivotal role that cannot be neglected~\cite{lowen2020inertial}.
For this reason, recent experimental and theoretical studies focus on the role of translational inertia in systems of active particles~\cite{fily2017mechanical, cecconi2018anomalous, lanoiselee2018statistical, caprini2021collective, te2021jerky, manacorda2017lattice, de2020phase, arold2020active, breoni2020active, vrugt2022microscopic} outlining how the usual scenario of overdamped active systems is modified. 
The single-particle properties are enriched because of additional transient regimes in the mean-square displacement~\cite{caprini2020inertial, nguyen2021active} as well as for the presence of a hidden entropy production~\cite{shankar2018hidden, crosato2019irreversibility, goswami2022inertial}. 
Inertia also affects both the virial stress and swim pressure~\cite{joyeux2016pressure, takatori2017inertial, gutierrez2020inertial} and modifies the transport properties in active density waves~\cite{zhu2018transport} and ratchet potentials~\cite{ai2017transport}, leading also to inhomogeneity and flux in the presence of a magnetic field~\cite{vuijk2020lorentz}. 

However, the non-equilibrium phase coexistence typical of active matter, known as motility induced phase separation~\cite{fily2012athermal, buttinoni2013dynamical, cates2015motility, digregorio2018full, martin2021characterization, caprini2020spontaneous}, is suppressed by translational inertia~\cite{mandal2019motility, dai2020phase, su2021inertia, omar2021tuning}.
Similarly, inertial effects reduce the accumulation near boundaries or obstacles typical of active particles~\cite{maggi2015multidimensional, deblais2018boundaries, caprini2018active} and hinder the crystallization~\cite{liao2021inertial}.
In addition, they promote hexatic ordering~\cite{negro2022inertial} in homogeneous phases and, in general, reduce the spatial velocity correlations~\cite{caprini2021spatial} characterizing dense active systems~\cite{caprini2020spontaneous, henkes2020dense, flenner2016nonequilibrium} both in liquids~\cite{marconi2021hydrodynamics, szamel2021long} or solid states~\cite{caprini2020hidden, caprini2021spatial}.
From the theoretical side, the main conclusion is that translational inertia mainly reduces the typical macroscopic properties characterizing active particles, apparently leading to "less active" systems more similar to their passive counterparts.

In spite of its important role in macroscopic experiments with inertial active particles~\cite{scholz2018inertial}, only recently the role of the rotational inertia, i.e. the effect of the moment of inertia, has been investigated. 
This additional ingredient has been introduced in Ref.~\cite{scholz2018inertial} to reproduce the inertial delay between particle orientation and velocity, experimentally observed in the behavior of a single active granular particle.
Successively, the problem has been further addressed theoretically~\cite{sprenger2021time, sandoval2020pressure} also through the introduction of a simplified model~\cite{lisin2022motion}.
Except for these contributions in the case of a single-particle, the role of rotational inertia on the collective phenomena typical of active particles has not been systematically explored and it will be the main object of investigation in this paper.

The article is structured as follows: in Sec.~\ref{sec:model} we introduce the model while in Sec.~\ref{sec:results} we show the numerical and theoretical results on the role of rotational inertia for collective phenomena, such as motility induced phase separation and spatial velocity correlations, as well as for the emergent coupling between rotational and translational velocities. 
Finally, we conclude in section~\ref{sec:conclusions}.


\section{Model}{\label{sec:model}}

\begin{figure*}[!t]
\centering
\includegraphics[width=0.9\linewidth,keepaspectratio]{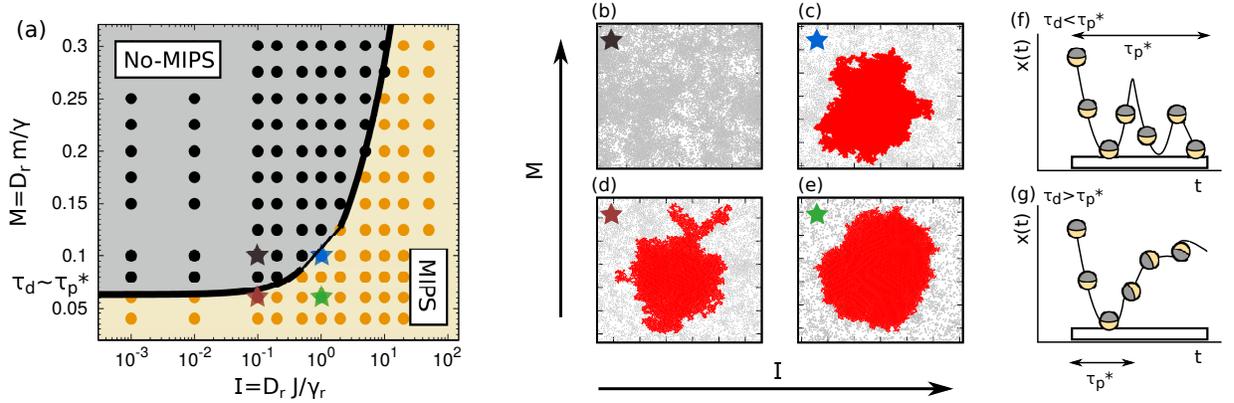}
\caption{Phase diagram. Panel (a): phase diagram: reduced mass $M=D_r m/\gamma$ (product between inertial time and rotational diffusion coefficient) vs reduced moment of inertia $\mathtt{I}=D_r J/\gamma_r$ (product between rotational time and rotational diffusion coefficient). The yellow region of the phase diagram corresponds to MIPS configurations while the grey regions to no-MIPS configurations characterized by an almost homogeneous phase.
The solid black line has been obtained by the scaling $\tau_d=m/\gamma \sim \tau_p^*$, where $\tau_p^*$ scales with $J/\gamma_r$ by Eq.\eqref{eq:tau_p*}.
The colored stars are drawn in correspondence with the parameters used to plot the snapshots of the configurations in panels (b), (c), (d), and (e), where the particles outside and inside the cluster are colored in grey and red, respectively, to clearly distinguish the boundary of the cluster.
Panels (f) and (g): schematic illustrations of the competition between bouncing effect and persistence time leading to the suppression of MIPS, showing the case $\tau_d \gg\tau_p^*$ (panel (f)) and $\tau_d \ll\tau_p^*$ (panel (g)). Simulations are realized with $N=10^4$.
}\label{fig:phasediagram}
\end{figure*}

We study a system of $N$ active Brownian particles (ABP)~\cite{fily2012athermal, cates2013active, farage2015effective, digregorio2018full, paoluzzi2022motility, caprini2022parental}, in a box of size $L$ with periodic boundary conditions. 
Each ABP with mass $m$ and moment of inertia $J$ evolves with an underdamped dynamics for the translational and orientational degrees of freedom, such that a particle with position, $\mathbf{x}_i$ and orientation $\theta_i$, is described by the following equations for the velocity $\mathbf{v}_i=\dot{\mathbf{x}}_i$ and the angular velocity $\omega_i=\dot{\theta}_i$
\begin{subequations}
\label{eq:activedynamics}
\begin{align}
&m\dot{\mathbf{v}}_i= -\gamma \mathbf{v}_i + \mathbf{F}_i +\sqrt{2 T \gamma} \boldsymbol{\xi}_i + \mathbf{f}^a_i\\
&J\dot{\omega}_i=-\gamma_r \omega_i + \gamma_r\sqrt{2D_r} \eta_i \,,
\end{align}
\end{subequations}
where $\boldsymbol{\xi}_i$ and $\eta_i$ are white noises with zero average and unit variance.
The coefficients $\gamma$ and $T$ are the friction coefficient and the temperature of the solvent bath, respectively, while $D_r$ and $\gamma_r$ are the rotational diffusion coefficient and the rotational friction coefficient.
The term $\mathbf{f}^a_i=\gamma v_0 \mathbf{n}_i$ describes the self-propulsion force which guarantees the persistence of the single-particle trajectory, $v_0$ being the swim velocity and $\mathbf{n}_i=(\cos\theta_i, \sin\theta_i)$ the orientational unit vector.
The particles interact through the force $\mathbf{F}_i=-\nabla_i U_{\text{tot}}$ due to a soft repulsive potential, $U_{\text{tot}}=\sum_{i<j} U(|\mathbf{x}_{i}-\mathbf{x}_{j}|)$, where $U=4\epsilon [(\sigma/r)^{12}-(\sigma/r)^{6}]$ is the WCA potential, with energy scale $\epsilon$ and particle diameter $\sigma$.
The ratios $\tau_d=m/\gamma$ and $\tau_r=J/\gamma_r$ define the typical translational and rotational inertial times, respectively, while $\tau_p=1/D_r$ corresponds to the persistent time of a single-trajectory.

Rescaling the position in units of $\sigma$ and the time in units of $\tau_p=1/D_r$, the system is controlled by many dimensionless parameters. A key role in active matter systems is provided by the P\'eclet number $\text{Pe}=v_0/(D_r\sigma)$, namely the ratio between persistence length, $v_0/D_r$, and particle diameter.  
The presence of the thermal bath introduces an additional dimensionless parameter, $\sqrt{T \gamma}/(D_r^{3/2}\sigma m)$, that induces the suppression of MIPS when random thermal fluctuations overcome the value of the active force. For this reason, we fix $\text{Pe}=50$ and $\sqrt{T \gamma}/(D_r^{3/2}\sigma m)=10^{-3}$ so that the effect of the thermal temperature is practically negligible with respect to that of the active force.
Recently, the influence of the potential details contained in the dimensionless parameter $\sqrt{\epsilon/m}/(D_r\sigma)$ has been investigated~\cite{de2022effect, martin2021characterization} and shows a shift in the MIPS transition. 

The presence of translational inertia has been recasted onto the reduced mass, 
\begin{equation}
M=\frac{D_r m}{\gamma} \,,
\end{equation}
defined as the ratio between the inertial time and the persistence time.
The rotational inertia gives rise to an additional dimensionless parameter that represents a reduced moment of inertia and reads
\begin{equation}
\mathtt{I}=\frac{D_r J}{\gamma_r} \,.
\end{equation}
This parameter induces non-trivial consequences in the behavior of a single active particle, such as the effect known as inertial delay~\cite{scholz2018inertial, sprenger2021time}, but could induce non-trivial dynamical consequences on the collective phenomena typical of active particles.
To evaluate the effect of the inertia, we only focus on $M$ and $\mathtt{I}$.

\section{Results}{\label{sec:results}}
\subsection{Phase diagram}

\begin{figure*}[!t]
\centering
\includegraphics[width=1\linewidth,keepaspectratio]{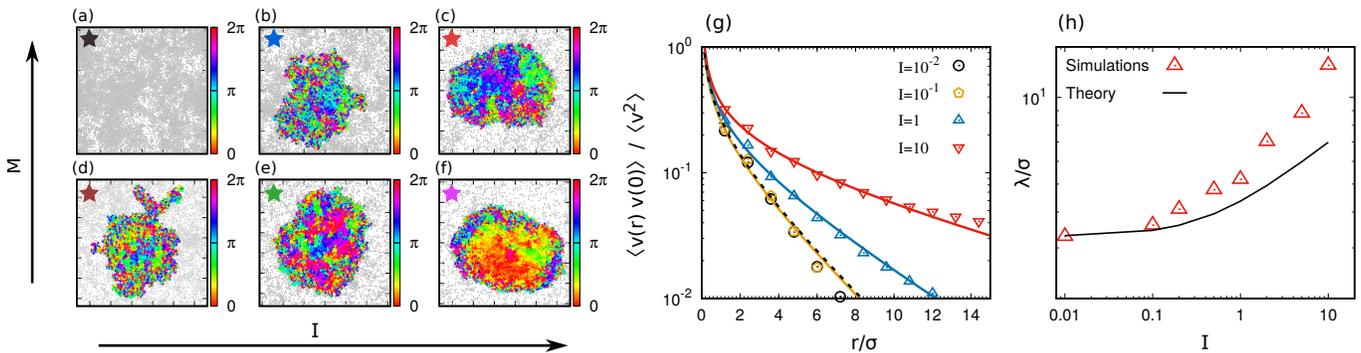}
\caption{
Spatial velocity correlations. Panels~(a)-(f): snapshot configurations for different values of $M=D_r m/\gamma$ and $\mathtt{I}=D_r J/\gamma_r$. Particles outside the cluster are colored in grey, while those inside the cluster are colored according to the angle formed by the velocity vector $\mathbf{v}_i$ and a reference axis $x$. These panels are obtained with $M=5\times10^{-2}, 10^{-1}$ (from below) and $\mathtt{I}=10^{-1}, 1, 10$ (from the left), $\text{Pe}=50$ and $N=10^4$.
Panel~(g): spatial velocity correlation, $\langle \mathbf{v}(r)\cdot\mathbf{v}(0)\rangle/\langle \mathbf{v}^2 \rangle$, as a function of $r/\sigma$, for different values of $\mathtt{I}=D_r J/\gamma_r$. Colored points are obtained by simulations while solid lines (colored accordingly) corresponds to the theoretical prediction, Eq.~\eqref{eq:spatialvelocitycorrelation}.
Panel~(h): correlation length, $\lambda$, as a function of $\mathtt{I}$. Points come from the numerical simulations and the solid black lines have been obtained implementing the theoretical prediction, Eq.~\eqref{eq:correlationlength}.
Simulations in panels (g) and (h) have been obtained with $M=5\times 10^{-2}$, $\text{Pe}=50$ and $N=10^4$.
}\label{Spatialvelocitycorrelation}
\end{figure*}

The impact of the rotational inertia is at first investigated by constructing the phase diagram in the plane spanned by reduced mass $M=D_r m/\gamma$ and reduced moment of inertia $\mathtt{I}=D_r J/\gamma_r$ (Fig.~\ref{fig:phasediagram}~(a)), for a packing fraction $\phi=N \sigma^2\pi/(4L^2)=0.5$.
For $\mathtt{I}\to0$ ($\mathtt{I} \ll 10^{-2}$), the results by Mandal et. al.~\cite{mandal2019motility} are quantitatively recovered: the increase of $M$ suppresses the coexistence of a dilute and a dense phase typical of pure repulsive active systems, known as motility induced phase separation (MIPS)~\cite{klamser2018thermodynamic, petrelli2020effective, maggi2021universality, caprini2020hidden, shi2020self}, (Fig.~\ref{fig:phasediagram}~(d)) and promotes an almost homogeneous phase (Fig.~\ref{fig:phasediagram}~(b)).
The increase of the rotational inertia favors MIPS, playing the opposite role than the translational inertia ($M$): the increase of reduced moment of inertia $\mathtt{I}$ drastically enhances the stability of the phase coexistence which can be achieved also for values of the reduced mass $M$ an order of magnitude larger than the case $\mathtt{I}\to0$ (See also the sequence of snapshots (b)$\to$(c) and (d)$\to$(e) in Fig.~\ref{fig:phasediagram}).
In short, rotational inertia induces MIPS.

This qualitative scenario can be understood in terms of intuitive dynamical scaling arguments, originally sketched to explain the formation of MIPS in overdamped active systems~\cite{fily2012athermal, redner2013structure, lowen2018active}. The persistence of the single-particle trajectory allows a couple of active particles to slow down (or stop) when they collide until the rotational diffusion changes the direction of their active forces.
When the average time between two collisions, $\tau_c$, is smaller than the typical persistence time, $\tau_p$, the condition for the cluster nucleation is achieved.
The rotational inertia provides an exponential memory to the particle orientation, hindering its change and, in first approximation, increasing the effective persistence time, $\tau_p = 1/D_r \to \tau_p^*$, from the overdamped value $1/D_r$ to a larger value depending on $J/\gamma_r$.
Following Ref.~\cite{sprenger2021time}, the analytical expression for $\tau_p^*$ can be analytically calculated as
\begin{equation}
\label{eq:tau_p*}
\tau_p^*\sim \frac{1}{D_r}
\begin{cases}
\left(1+D_r \frac{J}{\gamma_r}\right), \qquad D_r\frac{J}{\gamma_r} \ll 1 \,,\\
\sqrt{J /\gamma_r}, \qquad\qquad D_r\frac{J}{\gamma_r} \gg 1 \,,
\end{cases}
\end{equation}
by expanding the exact expression for $\tau_p^*$ in powers of $\mathtt{I}=D_r\frac{J}{\gamma_r}\ll1$ and $\mathtt{I}=D_r\frac{J}{\gamma_r}\gg 1$, respectively~\cite{sprenger2021time}. 

We remind that the increase of the translational inertia ($M$) suppresses MIPS because inertia induces bouncing effects that hinder the particle's ability to remain stuck in the cluster~\cite{mandal2019motility}.
The physical picture behind this mechanism is sketched in Fig.~\ref{fig:phasediagram}~(f) and~(g).
A particle approaching the cluster persistently bounces on the cluster surface if its persistence time is larger than its typical bouncing time, i.e. the time occurring between two successive bounces (panel~(f)).
When this condition is achieved and the collisional time is larger than the persistence time, the cluster forms.
If the active particle reorients during the first bouncing event, the particle effectively behaves as a passive particle (panel~(g)) and will never remain stuck in the cluster, or in other words, the cluster cannot nucleate.
Taking in mind this picture and that the bouncing time is roughly proportional to the inertial time $\tau_d$, we predict analytically the scaling of the transition line between MIPS and no-MIPS configurations. Since the rotational inertia leads to the effective persistence time $\tau_p^*$, such a scaling is achieved by simply requiring that the inertial time equals the effective persistent time
\begin{equation}
\tau_d= m/\gamma \sim \tau_p^* \,.
\end{equation}
Through this analytical argument, we recover the scaling of the transition line with the rotational inertia, namely $M \sim (1+ D_r J/\gamma_r)$ and $M\sim \sqrt{J/\gamma_r}$ for small and large $J/\gamma_r$, respectively (see solid black lines in Fig.~\ref{fig:phasediagram}~(a)).

\subsection{Rotational inertia promotes velocity order}

The sequences of snapshots (a)$\to$(b)$\to$(c) and (d)$\to$(e)$\to$(f) in Fig.~\ref{Spatialvelocitycorrelation} suggest that the increase of the rotational inertia promotes also the more recent collective phenomenon observed in active Brownian particles: the spontaneous emergence of velocity alignment and spatial velocity correlations~\cite{caprini2020spontaneous, caprini2020time}.
The particles in the dense cluster are colored according to their velocity orientation, say the angle formed by the particle velocity $\mathbf{v}_i$ and a reference axis (say $x$), and show the emergence of regions where the particles move in the same direction, despite the absence of any alignment interaction.
The larger is the reduced moment of inertia $\mathtt{I}$, the larger the size of the region with the same color, until for the larger values of $\mathtt{I}$ (for instance $\mathtt{I}=10$ and $M=0.05$, Fig.~\ref{fig:phasediagram}~(g)), velocity domains have a size comparable with the cluster size.

Spatial velocity correlations, $\langle \mathbf{v}(r)\cdot\mathbf{v}(0)\rangle/\langle \mathbf{v}^2 \rangle$, are investigated to quantify this qualitative picture and are reported in Fig.~\ref{Spatialvelocitycorrelation}~(g).
They show an exponential-like behavior with a typical correlation length, $\lambda$, that increases as $\mathtt{I}=D_rJ/\gamma_r$ becomes larger, as revealed by Fig.~\ref{Spatialvelocitycorrelation}~(h).
In agreement with previous theoretical works originally developed in the context of active particles in solid configurations~\cite{caprini2020hidden, caprini2021spatial}, the spatial profile of $\langle \mathbf{v}(r)\cdot\mathbf{v}(0)\rangle$ calculated in the bulk of the dense cluster reads (see solid lines in Fig.~\ref{Spatialvelocitycorrelation}~(g)):
\begin{equation}
\label{eq:spatialvelocitycorrelation}
\langle \mathbf{v}(r)\cdot\mathbf{v}(0)\rangle \sim \frac{e^{-r /\lambda}}{r^{1/2}} \,,
\end{equation}
where $\lambda$ represents the correlation length of the spatial velocity correlation 
\begin{equation}
\label{eq:correlationlength}
\lambda \propto \frac{3}{2} \frac{(\tau_p^*)^2}{1+\frac{\tau_p^*}{\tau_d}} v_s \,.
\end{equation}
In the case of active liquids, the parameter $v_s$ contains the main dependence on the density and swim velocity $v_0$, being proportional to the bulk modulus (see Refs.~\cite{szamel2021long, marconi2021hydrodynamics} for details), while in the case of active solids is purely determined by the local packing fraction and completely independent of $v_0$~\cite{caprini2020hidden, caprini2021spatial}.
With respect to the original prediction, here, the expression for $\lambda$ is modified by replacing $\tau_p \to \tau_p^*$ to account for the effective persistence time induced by the rotational inertia.

In Fig.~\ref{Spatialvelocitycorrelation}~(h), the Eq.~\eqref{eq:correlationlength} is compared with the value of $\lambda$ obtained by numerical simulations.
Our prediction~\eqref{eq:correlationlength}, obtained by fixing $v_s$ to the case without rotational inertia, qualitatively reproduces the increase of $\lambda$ with the reduced moment of inertia $\mathtt{I}$ but fails quantitatively because it underestimates its value.
This occurs because, for phase-separated configurations, the increase of rotational inertia increases the local packing fraction of the cluster, and, thus, the value of $v_s$.



\subsection{Coupling between rotational and translational velocity}

\begin{figure}[!t]
\centering
\includegraphics[width=1\linewidth,keepaspectratio]{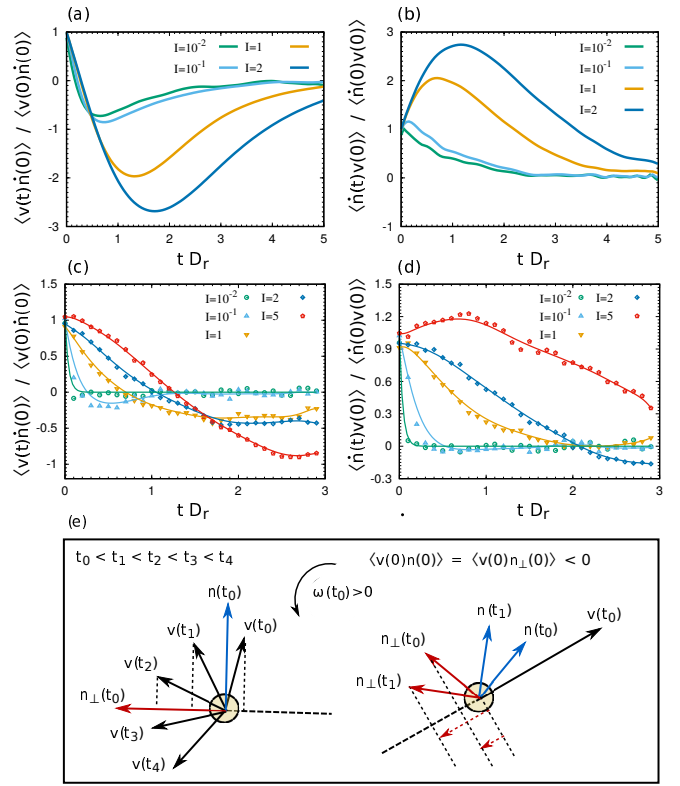}
\caption{
Coupling between translational and rotational inertia.
Panels~(a) and~(c): autocorrelation $\langle \mathbf{v}(t)\cdot \dot{\mathbf{n}}(0)\rangle$  for different values of the rotational inertia, through the reduced moment of inertia $\mathtt{I}$, for a potential-free particle (panel~(a)) and a homogeneous dense phase (panel~(c)).
Panels~(b) and~(d): autocorrelation $\langle \dot{\mathbf{n}}(t)\cdot \mathbf{v}(0)\rangle$ for different values of the rotational inertia $\mathtt{I}$ for a potential-free particle (panel~(b)) and a homogeneous dense phase (panel~(d)).
Panel~(e): schematic time evolution (when the rotational inertia is relevant) of the velocity vector $\mathbf{v}(t)$ (black), the self-propulsion vector $\mathbf{n}(t)$ (blue) and its normal vector on the plane of motion $\mathbf{n}_\perp(t)$ (red). 
For a reference case, we have fixed $\omega(t)>0$ so that $\mathbf{n}(t)$ and $\mathbf{n}_\perp(t)$ rotate anti-clockwise.
}\label{fig:fig3}
\end{figure}

Despite the first-order qualitative picture of effective persistence time helps to understand the main collective phenomena of active particles, the increase of the rotational inertia leads also to peculiar properties absent in the case $\mathtt{I}\to 0$.
We focus on the autocorrelation between translational and rotational velocities, say $\langle \mathbf{v}(t)\cdot \dot{\mathbf{n}}(0)\rangle$ and its time-reversed $\langle \dot{\mathbf{n}}(t)\cdot \mathbf{v}(0)\rangle$.
These observables are plotted in Fig.~\ref{fig:fig3} in two main cases: i) non-interacting particles, which approximate the dilute regions outside the cluster (panels~(a) and~(b)) and ii) a dense homogeneous case, with packing fraction $\phi$ of the same order of the packing fraction reached in a typical cluster phase (panels~(c) and~(d)).
For $\mathtt{I} \to 0$, the angular velocity evolves on a time-scale faster than the one characterizing the evolution of the translational velocity and 
as a consequence, $\langle \mathbf{v}(t)\cdot \dot{\mathbf{n}}(0)\rangle/\langle \mathbf{v}(0)\cdot \dot{\mathbf{n}}(0)\rangle$ and $\langle \dot{\mathbf{n}}(t)\cdot \mathbf{v}(0)\rangle/\langle \mathbf{v}(0)\cdot \dot{\mathbf{n}}(0)\rangle$ display time-profiles fast decreasing towards zero.
The two normalized time correlations display more complex temporal profiles when the reduced moment of inertia $\mathtt{I}$ is increased.
$\langle \dot{\mathbf{n}}(t)\cdot {\mathbf{v}}(0)\rangle/\langle \mathbf{v}(0)\cdot \dot{\mathbf{n}}(0)\rangle$ grows in time in a first time-regime and then decreases slower towards zero as $\mathtt{I}$ is increased (panels (b) and (d)).
On the contrary, $\langle \mathbf{v}(t)\cdot \dot{\mathbf{n}}(0)\rangle/\langle \mathbf{v}(0)\cdot \dot{\mathbf{n}}(0)\rangle$ decreases and reaches negative values (smaller as $\mathtt{I}$ is increased), approaches a minimum, and then grows until zero (panels~(a) and~(c)).
We also point out that in the denser cases both autocorrelations display a time-profile less pronounced with respect to the dilute cases.
This occurs because the strong interactions hinder the ability of the particle velocity to align to the self-propulsion vector by producing an effectively randomizing effect that reduces the cross-correlations.

In systems of active matter, rotational and translational degrees of freedom mix, and, in particular, the former affects the latter.
This result has not an equilibrium counterpart: indeed, for $v_0 \to 0$ the equation of motion for $\mathbf{v}$ is not coupled to the one for $\mathbf{n}$ and, thus, $\langle \dot{\mathbf{n}}(t)\cdot {\mathbf{v}}(0)\rangle = 0$ and $\langle \mathbf{v}(t)\cdot \dot{\mathbf{n}}(0)\rangle=0$ for all times $t$.

The shape displayed by the autocorrelations for large $\mathtt{I}$ has an intuitive explanation because we can express
$\dot{\mathbf{n}}(t) = \omega(t) \mathbf{z} \times \mathbf{n}(t)= \omega(t)\mathbf{n}_{\perp}(t)$, where $\mathbf{z}$ is the axis normal to the plane of motion and $\mathbf{n}_{\perp}$ is the vector orthogonal to $\mathbf{n}$ again on the plane of motion.
As a consequence, we can express $\langle \mathbf{v}(t)\cdot \dot{\mathbf{n}}(0)\rangle = \langle \mathbf{v}(t)\cdot \mathbf{n}_{\perp}(0) \omega(0)\rangle$ and $\langle \mathbf{v}(0)\cdot \dot{\mathbf{n}}(t)\rangle = \langle \mathbf{v}(0)\cdot \mathbf{n}_{\perp}(t) \omega(t)\rangle$.
A schematic representation of the scalar product involded in these time-correlations is reported in Fig.~\ref{fig:fig3}~(e), where the projections of $\mathbf{v}(t)$ over $\dot{\mathbf{n}}_\perp$ (and viceversa) are shown for different times.
By choosing $\omega(0)>0$ and assuming a large value of $\mathtt{I}$ such that $\omega(t)$ remains positive until time $t$, $\mathbf{n}$ and $\mathbf{n}_\perp$ rotates counter-clockwise.
The typical inertial delay due to the rotational inertia~\cite{scholz2018inertial, sprenger2021time} implies that at the initial time $t_0$ the velocity $\mathbf{v}(t_0)$ typically forms an angle $0<\alpha<\pi/2$ with $\mathbf{n}$. By increasing time, $\mathbf{v}(t)$ tends to align to $\mathbf{n}(t)$ and, thus, rotates counterclockwise.
As a consequence, at $t=t_0$ the projection of $\mathbf{v}(t_0)$ over $\mathbf{n}_\perp(t_0)$ is negative and thus, $\langle \mathbf{v}(t)\cdot \dot{\mathbf{n}}(0)\rangle<0$, while for successive times the projection of $\mathbf{v}(t_0)$ over $\mathbf{n}_\perp(t_0)$ becomes positive. This argument clearly explains the non-monotonic behavior observed in Fig.\ref{fig:fig3} (a) and (c).
Similarly, we can understand the increase of $\langle \mathbf{v}(0)\cdot \dot{\mathbf{n}}(t)\rangle$: the projection of $\mathbf{n}_\perp(t)$ over $\mathbf{v}(t_0)$ is always negative and becomes larger in modulus for the time $t>t_0$ when $\mathbf{n}_\perp(t)$ has the same direction of $\mathbf{v}(t_0)$ (we remind that $\langle \mathbf{v}(0)\cdot \dot{\mathbf{n}}(0)\rangle<0$).

\section{Discussion and Conclusions}{\label{sec:conclusions}}

In conclusion, we have explored the influence of rotational inertia on collective phenomena in active matter.
The standard scenario of motility induced phase separation (MIPS) is typically encountered for self-propelled spherical particles with overdamped Brownian dynamics~\cite{fily2012athermal, palacci2013living, fodor2016far, caporusso2020motility, caprini2020spontaneous}.
However, MIPS is disfavored in a broad range of cases that go beyond this simple setup.
Indeed, anisotropic rod-like particle shapes~\cite{moran2022particle, theers2018clustering, grossmann2020particle}, aligning interactions~\cite{van2019interrupted, sese2021phase, pu2017reentrant}, circling torques~\cite{liao2018clustering, ma2022dynamical}, hydrodynamic interactions~\cite{matas2014hydrodynamic, theers2018clustering}, polydispersity~\cite{paoluzzi2022motility, kumar2021effect} and translational inertia~\cite{mandal2019motility, dai2020phase, su2021inertia, omar2021tuning} typically act against MIPS.
Contrarily, we have shown here that rotational inertia favors MIPS by contributing to increasing the effective persistence time of the particle dynamics.
Concomitantly, this effect enhances the spatial velocity correlations displayed by dense active systems counteracting its suppression due to translational inertia~\cite{caprini2021spatial}.

We emphasize that this effect is experimentally verifiable for vibrated granulate particles where the mass and moment of inertia can systematically be changed and tuned.
As a consequence, our study suggests how to design granular active particles to enhance the typical collective phenomena displayed by microscopic active matter systems.
From a theoretical side, the generalization of approximated methods to the inertial case, ranging from a modified Maxwell constructions~\cite{solon2018generalized} to reproducing the phase diagram and equilibrium-like approaches to calculate the effective interactions between particles~\cite{farage2015effective, wittmann2017effective, wittmann2017effectiveII}, represents an interesting perspective that could stimulate further studies.

 \section*{Acknowledgements}
LC thanks Alexander Ralf Sprenger for illuminating discussions.
LC acknowledges support from the Alexander Von Humboldt foundation.
HL acknowledge support by the Deutsche Forschungsgemeinschaft (DFG) through the SPP 2265 under the grant number LO 418/25-1.



\bibliographystyle{apsrev4-1}

\bibliography{under.bib}

\end{document}